\documentclass[showpacs,showkeys,11pt,
preprint,preprintnumbers,nofootinbib,
groupedaddress,superscriptaddress,amsmath,amssymb]{revtex4}
\usepackage{amsfonts}
\usepackage{graphics}
\usepackage{epsfig}
\usepackage{url}
\usepackage{multirow}
\usepackage{feynmp}
\newcommand {\be}{\begin{equation}}
\newcommand {\ee}{\end{equation}}
\newcommand {\ba}{\begin{eqnarray}}
\newcommand {\ea}{\end{eqnarray}}

\newcommand {\tanb}{$\tan\beta~$}
\newcommand {\ra}{\rightarrow}

\begin{document}
\title{Charged Higgs Pair Production in a General Two Higgs Doublet Model at $e^+e^-$ and $\mu^+\mu^-$ Linear Colliders}
%%%%%%%%%%%%%%%%%%%%%%%%%%%%%%%%%%%%%%%%%%%%%%%%%%%%%%%%%%%%%%%%%%%%%%%%%%%%%%%%
\author{M. Hashemi}
\email{hashemi_mj@shirazu.ac.ir}
\affiliation{Physics Department and Biruni Observatory, College of Sciences, Shiraz University, Shiraz 71454, Iran}
%%%%%%%%%%%%%%%%%%%%%%%%%%%%%%%%%%%%%%%%%%%%%%%%%%%%%%%%%%%%%%%%%%%%%%%%%%%%%%%%
%\date{\today}

\begin{abstract}
In this paper, charged Higgs pair production through $\ell^+ \ell^- \ra H^+ H^-$ where $\ell = e~ \textnormal{or} ~\mu$, is studied within the framework of a general Two Higgs Doublet Model (2HDM). The analysis is relevant to a future $e^+e^-$ or $\mu^+\mu^-$ collider operating at center of mass energy of $\sqrt{s}=500$ GeV. Two different scenarios of small and large $\alpha$ values is studied. Here $\alpha$ is the parameter which diagonalizes the neutral CP-even Higgs boson mass matrix. Within the Minimal Supersymmetric Standard Model (MSSM), cross section of this process is almost the same at $e^+e^-$ and $\mu^+\mu^-$ colliders. It is shown that at $e^+e^-$ colliders within a general 2HDM, cross section is not sensitive to the mass of neutral Higgs bosons, however, it can acquire large values up to several picobarn at $\mu^+\mu^-$ colliders with the presence of heavy neutral Higgs bosons. A scan over Higgs boson mass parameter space is performed to analyze the effect of large masses of neutral Higgs bosons involved in the s-channel propagator and thus in the total cross section of this process. 
\end{abstract}

\maketitle

\section{Introduction}
The Standard Model of Particle Physics (SM) has been tested experimentally during the last decades and a beautiful agreement between the theoretical predictions and experimental data has been achieved. The Higgs mechanism is assumed to be the right approach for giving masses to the massless electroweak particles and gauge bosons \cite{Higgs1,Higgs2,Higgs3,Higgs4,Higgs5}. A signal has already been observed with a mass around 125 GeV, and is believed to be the SM Higgs boson predicted by the Higgs mechanism \cite{125atlas,125cms}. However, SM with one single Higgs boson suffers from shortcomings which motivate theoretical extentions. One of those is the Higgs boson mass quadratic divergence when radiative corrections are included. One of solutions to this problem is to introduce Supersymmetry \cite{susy,susy2} which requires a non-minimal Higgs sector. The Minimal Supersymmetric Standard Model (MSSM) is the simplest example of a supersymmetric model which belongs to two Higgs doublet models (2HDM), i.e., it requires two Higgs doublets to give masses to leptons and quarks \cite{Martin}. In a general 2HDM, two charged Higgs bosons, $H^{\pm}$, two CP-event neutral Higgs bosons, $h^0,~H^0$, and a CP-odd Higgs, $A^0$, are predicted. The lightest neutral Higgs boson, $h^0$, is taken to be SM-like and is the candidate for the signal observed at LHC. The rest have escaped from detection so far. The main reasons could be the lower production rate or difficulty in extracting the signal from the SM background. In this paper the focus is on the charged Higgs boson which provides a unique and different signal due to being charged. There has been a long time attempt to observe a signal associated with this particle in the last and current experiments. Results from LEP exclude a charged Higgs with $m_{H^{\pm}}<89~ \textnormal{GeV}$ for all \tanb values \cite{lepexclusion1}. The Tevatron searches by D0 \cite{d01,d02,d03,d04} and CDF \cite{cdf1,cdf2,cdf3,cdf4} allow $2 <$ tan$\beta < 30$ for $m(H^{\pm}) > 80 $ GeV. The available area in terms of \tanb is larger for heavier charged Higgs bosons. The current direct search results from LHC exclude a charged Higgs boson with $m_{H^{\pm}}=90$ GeV if \tanb $>10$ \cite{atlasdirect,cmsdirect}. The limit on \tanb is weaker for heavier charged Higgs bosons. For instance a charged Higgs boson with $m_{H^{\pm}}=150$ GeV may exist with \tanb $<50$. These limits are obtained within MSSM. There are limits from B-Physics experiments which are stronger than those obtained from direct searches. The CLEO collaboration excludes a charged Higgs mass below 300 GeV at 95 $\%$ C.L. in 2HDM Type II with \tanb higher than 2 \cite{B1}. A study of four types of 2HDMs excludes $m_{H^{\pm}}~<~300$ GeV in 2HDM Type II and III while types I and IV have no lower limit for the charged Higgs mass at high \tanb \cite{main2hdm}. \\
\section{Theoretical Framework}
The most general potential using two Higgs doublets can be written in the form \cite{2hdm1,2hdm2}
\begin{align}
\mathcal{V} = & m_{11}^2\Phi_{1}^{\dag}\Phi_{1} + m_{22}^2\Phi_{2}^{\dag}\Phi_{2} - \left[m_{12}^2\Phi_{1}^{\dag}\Phi_{2}+\textnormal{h.c.}\right] \notag \\
&+\frac{1}{2}\lambda_1\left(\Phi_{1}^{\dag}\Phi_{1}\right)^2+\frac{1}{2}\lambda_2\left(\Phi_{2}^{\dag}\Phi_{2}\right)^2
+\lambda_3\left(\Phi_{1}^{\dag}\Phi_{1}\right)\left(\Phi_{2}^{\dag}\Phi_{2}\right)
+\lambda_4\left(\Phi_{1}^{\dag}\Phi_{2}\right)\left(\Phi_{2}^{\dag}\Phi_{1}\right) \notag \\
&+\left\lbrace\frac{1}{2}\lambda_5\left(\Phi_{1}^{\dag}\Phi_{2}\right)^2+\left[\lambda_6\left(\Phi_{1}^{\dag}\Phi_{1}\right)
+\lambda_7\left(\Phi_{2}^{\dag}\Phi_{2}\right)\right]\left(\Phi_{1}^{\dag}\Phi_{2}\right)+\textnormal{h.c.}\right\rbrace 
\end{align} 
The free parameters are taken usually as 
\be
\lambda_1,\lambda_2,\lambda_3,\lambda_4,\lambda_5,\lambda_6,\lambda_7,m_{12}^{2},\beta 
\ee
in the general basis. The CP violation and Flavor Changing Neutral Currents (FCNC) are naturally suppressed via the Natural Flavor Conservation (NFC) mechanism if the $Z_{2}$ symmetry is imposed on the Lagrangian \cite{weinberg}. The $Z_{2}$ symmetry is defined as $\Phi_{i}\rightarrow(-1)^{i}\Phi_{i}$ ($i=1,2$) which has been discussed in details in \cite{2hdm3}. As a consequence one arrives at the following requirement,
\be
\lambda_6=\lambda_7=0
\ee
which should be respected in CP-conserving models. The values of $\lambda_i$ can be expressed in terms of the Higgs boson masses, $m_{12}^2$, $\alpha$, $\beta$, $\lambda_6$ and $\lambda_7$ \cite{2hdm4}. Therefore in a CP-conserving 2HDM, the set of free parameters can be taken as 
\be
m_h,m_H,m_A,m_{H^{\pm}},\alpha,\beta,m_{12}^2.
\ee
A characteristic feature of SUSY models is that $\lambda_5=0$ \cite{HHH,HHH2}. This requirement has already been applied in MSSM \cite{mm1,mm2,mm3}, i.e.,  
\be
\lambda_5=\lambda_6=\lambda_7=0.
\ee
In a general 2HDM, the above setting symplifies the set of parameters as it can be used to express $m_{12}^2$ in terms of $m_{A}$ through the following 2HDM relation,
\be
m_A^2=\frac{m_{12}^2}{\sin\beta\cos\beta}-\frac{v^2}{2}(2\lambda_5+\lambda_6\cot\beta+\lambda_7\tan\beta).
\ee
which reduces to $m_{12}^2=m_A^2 \cos\beta \sin\beta$ if $\lambda_5=\lambda_6=\lambda_7=0$. Therefore one can choose the following subset of parameters to describe the model,
\be
m_h,m_H,m_A,m_{H^{\pm}},\alpha,\beta.
\ee
or equivalently 
\be
m_h,m_H,m_A,m_{H^{\pm}},\cos(\beta-\alpha),\tan\beta.
\ee
Throughout this paper, we adopt this setting and let the Higgs bosons masses be free while checking the general potential in terms of stability (positivity) and unitarity with the use of 2HDMC 1.1 \cite{2hdmc}.\\
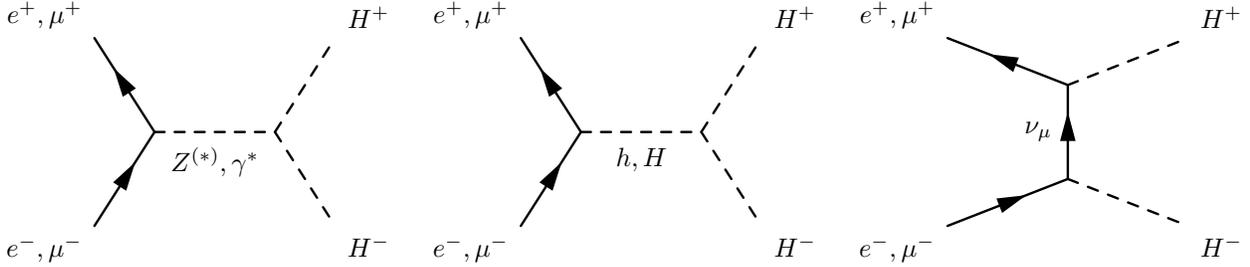
\begin{figure}
\begin{minipage}[t]{0.3\linewidth}
\centering
\unitlength=1mm
\begin{fmffile}{schannel}
\begin{fmfgraph*}(40,25)
\fmfleft{i1,i2}
\fmfright{o1,o2}
\fmflabel{$e^-,\mu^-$}{i1}
\fmflabel{$e^+,\mu^+$}{i2}
\fmflabel{$H^+$}{o2}
\fmflabel{$H^-$}{o1}
\fmf{fermion}{i1,v1,i2}
\fmf{dashes}{v2,o2}
\fmf{dashes}{v2,o1}
\fmf{photon,label=$Z^{(*)},,\gamma^{*}$}{v1,v2}
\end{fmfgraph*}
\end{fmffile}
\end{minipage}
\hspace{0.5cm}
\begin{minipage}[t]{0.3\linewidth}
\centering
\unitlength=1mm
\begin{fmffile}{schannel2}
\begin{fmfgraph*}(40,25)
\fmfleft{i1,i2}
\fmfright{o1,o2}
\fmflabel{$e^-,\mu^-$}{i1}
\fmflabel{$e^+,\mu^+$}{i2}
\fmflabel{$H^+$}{o2}
\fmflabel{$H^-$}{o1}
\fmf{fermion}{i1,v1,i2}
\fmf{dashes}{v2,o2}
\fmf{dashes}{v2,o1}
\fmf{dashes,label=$h,,H$}{v1,v2}
\end{fmfgraph*}
\end{fmffile}
%\label{a}
\end{minipage}
\hspace{0.5cm}
\begin{minipage}[t]{0.3\linewidth}
\centering
\unitlength=1mm
\begin{fmffile}{tchannel}
\begin{fmfgraph*}(40,25)
\fmfleft{i1,i2}
\fmfright{o1,o2}
\fmflabel{$\mu^-$}{i1}
\fmflabel{$\mu^+$}{i2}
\fmflabel{$H^+$}{o2}
\fmflabel{$H^-$}{o1}
\fmf{fermion}{i1,v1}
\fmf{fermion}{v2,i2}
\fmf{fermion,label=$\nu_{\mu}$}{v1,v2}
\fmf{dashes}{v2,o2}
\fmf{dashes}{v1,o1}
\end{fmfgraph*}
\end{fmffile}
%\label{b}
\end{minipage}
\caption{The $s-$channel (left and middle) and $t-$channel (right) diagrams involved in the signal process. \label{diagrams}}
\end{figure}
\section{The Charged Higgs Pair Production at $e^-e^+$ Linear Colliders}
A linear $e^{-}e^{+}$ collider operating at a center of mass energy of 500 GeV has a large potential for a charged Higgs observation through the pair production process $e^+e^- \ra H^+H^-$. Such a collider may be the International Linear Collider (ILC) \cite{ilc,rdr} or the Compact Linear Collider (CLIC) \cite{clic} operating in its low energy phase. \\
The charged Higgs pair production, $e^{+}e^{-}\ra H^{+}H^{-}$, has been studied in \cite{pair,mine}. When including off-shell effects, the charged Higgs can be produced through $e^{+}e^{-}\ra \tau\bar{\nu} H^{+}$ \cite{sch6,sch7}. Results from \cite{sch6,sch7} show that including off-shell effects, the $5\sigma$ contour would be extended by about 10 GeV compared to the case of on-shell charged Higgs pair production. In this paper the focus is on an on-shell pair production of charged Higgs bosons, although including off-shell effects could increase the cross section near the kinematic threshold.\\
There has been a number of studies of charged Higgs production at $e^+e^-$ colliders \cite{sch1,sch2,sch3,sch4,sch5,sch8,WH1,WH2,WH3,WH4,WH5}. However, none of the production processes lead to a more promising result than the charged Higgs pair production followed by a decay to $t\bar{b}$ or $\tau\nu$.\\
\section{Cross Section at $e^+e^-$ and $\mu^+\mu^-$ Colliders}
The charged Higgs pair production proceeds through three types of Feynman diagrams shown on Fig. \ref{diagrams}. The left diagram includes s-channel electroweak propagators, while the middle one is an s-channel diagram consisting of neutral Higgs bosons propagators. The CP-odd $A^0$ does not contribute to the production process due to zero coupling with the charged Higgs pair. Therefore we only deal with CP-even neutral Higgs bosons, $h^0$ and $H^0$. The right diagram shows a t-channel process which is negligible at $e^+e^-$ and $\mu^+\mu^-$ colliders due to the low Yukawa coupling between the charged Higgs boson and the lepton ($e$ or $\mu$). Within the MSSM, the neutral Higgs bosons have a very small contribution to the total cross section of this process which is due to the fact that they are not so heavy to be produced as a resonance. On the contrary in a general 2HDM where the neutral Higgs boson masses are free, sizable contributions to the total cross section can be achieved. In order to illustrate this, two points A and B are selected with their coordinates as follows.
\be
\textnormal{Point A}: ~m_h=125~ \textnormal{GeV},~m_A=m_H=150 ~\textnormal{GeV},~m_{H^{\pm}}=170~\textnormal{GeV}
\ee
\be
\textnormal{Point B}: ~m_h=125~ \textnormal{GeV},~m_A=150 ~\textnormal{GeV},~m_H=350 ~\textnormal{GeV},~m_{H^{\pm}}=170~\textnormal{GeV}
\ee
The point A is within the MSSM phase space, while point B deviates from MSSM due to representing a heavy neutral Higgs ($H$). Therefore they are expected to show the effect of a heavy neutral Higgs boson ($H$) in the total cross section at two different types of collisions. A similar effect is expected to be observed when a heavy $h$ is involved. The value of $\alpha$ parameter in MSSM is obtained using FeynHiggs 2.8.3 \cite{fh1,fh2,fh3,fh4} which gives $\alpha=-0.4$ at 2-loop level calculation using point A as input. This value is also used for point B. The $\beta$ parameter is chosen to satisfy $\tan\beta=10$. Table \ref{points} shows cross sections obtained from each diagram (left and middle) of Fig. \ref{diagrams} individually and the total cross section at $e^+e^-$ and $\mu^+\mu^-$ colliders. \\
As is seen, at $\mu^+\mu^-$ colliders, where the Higgs-muon couplings are larger by a factor of 200 compared to Higgs-electron couplings at $e^+e^-$ colliders, the s-channel diagrams which involve neutral Higgs bosons start to be important when the neutral Higgs boson mass is increased. This conclusion is generally independent of the type of 2HDM or neutral-charged Higgs coupling and only reflects the effect of the larger Yukawa coupling between the neutral Higgs boson and leptons when electrons are replaced by muons. A detailed discussion on the type of 2HDM and Higgs self coupling is presented in the next sections. Since the $e^+e^-$ collider is almost insensitive to the mass of the neutral Higgs bosons, the rest of this study focuses on a $\mu^+\mu^-$ collider operating at $\sqrt{s}$ = 500 GeV.   
\begin{table}[h]
\begin{center}
\begin{tabular}{|c|c|c|c|c|c|c|}
\hline
Collider type& \multicolumn{3}{c|}{$e^+e^-$} & \multicolumn{3}{c|}{$\mu^+\mu^-$} \\
\hline
Channel & $Z/\gamma$ & $h^0/H^0$ & Total & $Z/\gamma$ & $h^0/H^0$ & Total  \\
\hline
Point A & 43$~fb$ & 5$\times 10^{-9}~fb$ & 43$~fb$ & 47$~fb$ & 2 $\times 10^{-4}~fb$ & 47$~fb$ \\
\hline
Point B & 43$~fb$ &  $10^{-3}~fb$ & 43$~fb$ & 47$~fb$ & 38$~fb$ & 85$~fb$ \\
\hline
\end{tabular}
\end{center}
\caption{Cross section of different production channels at $e^+e^-$ and $\mu^+\mu^-$ colliders. \label{points}}
\end{table}
\section{The type of 2HDM and the neutral Higgs coupling to fermions}
In this section, the neutral Higgs boson couplings to the muon pair and the charged Higgs pair are studied. First recall that the Higgs-fermion interaction Lagrangian can be written as follows \cite{yukawa}:
\begin{align}
-\mathcal{L}=&\frac{1}{\sqrt{2}}\bar{D}\left \lbrace\kappa^D s_{\beta-\alpha}+\rho^D c_{\beta-\alpha} \right \rbrace Dh
+\frac{1}{\sqrt{2}}\bar{D}\left \lbrace\kappa^D c_{\beta-\alpha}-\rho^D s_{\beta-\alpha} \right \rbrace DH + \frac{i}{\sqrt{2}}\bar{D}\gamma_5\rho^D DA \notag\\ 
&+\frac{1}{\sqrt{2}}\bar{U}\left \lbrace\kappa^U s_{\beta-\alpha}+\rho^U c_{\beta-\alpha} \right \rbrace Uh
+\frac{1}{\sqrt{2}}\bar{U}\left \lbrace\kappa^U c_{\beta-\alpha}-\rho^U s_{\beta-\alpha} \right \rbrace UH - \frac{i}{\sqrt{2}}\bar{U}\gamma_5\rho^U UA \notag\\
&+\frac{1}{\sqrt{2}}\bar{L}\left \lbrace\kappa^L s_{\beta-\alpha}+\rho^L c_{\beta-\alpha} \right \rbrace Lh
+\frac{1}{\sqrt{2}}\bar{L}\left \lbrace\kappa^L c_{\beta-\alpha}-\rho^L s_{\beta-\alpha} \right \rbrace LH + \frac{i}{\sqrt{2}}\bar{L}\gamma_5\rho^L LA \notag\\
&+\left[\bar{U}\left \lbrace V_{CKM}\rho^D P_R-\rho^U V_{CKM} P_L \right \rbrace DH^+ + \bar{\nu}\rho^LP_RLH^+ + \textnormal{h.c.} \right],
\label{Lagrangianeq}
\end{align}
where the following abbreviations have been used: $s_{\beta-\alpha}=\sin(\beta-\alpha)$, $c_{\beta-\alpha}=\cos(\beta-\alpha)$, $\rho^Q=\lambda^Q\kappa^Q$, $\kappa^Q=\sqrt{2}\frac{m^Q}{v}$. The $\lambda^Q$ parameter determines the type of the 2HDM \cite{types} according to Tab. \ref{tabtypes}.
\begin{table}[h]
\begin{tabular}{|c|c|c|c|c|}
\hline
\multicolumn{5}{|c|}{Type}\\
& I & II & III & IV \\
\hline
$\rho^D$ & $\kappa^D \cot\beta$ & $-\kappa^D \tan\beta$ & $-\kappa^D \tan\beta$ & $\kappa^D \cot\beta$ \\
\hline  
$\rho^U$ & $\kappa^U \cot\beta$ & $\kappa^U \cot\beta$ & $\kappa^U \cot\beta$ & $\kappa^U \cot\beta$ \\
\hline  
$\rho^L$ & $\kappa^L \cot\beta$ & $-\kappa^L \tan\beta$ & $\kappa^L \cot\beta$ & $-\kappa^L \tan\beta$ \\
\hline  
\end{tabular}
\caption{The four types of a general 2HDM in terms of the couplings in the Higgs-fermion Yukawa sector. \label{tabtypes}}
\end{table}
In a general 2HDM, the neutral Higgs coupling to $\mu^+\mu^-$ is obtained from Eq. \ref{Lagrangianeq} as follows:
\be
\textnormal{Type II, IV}:~\bar{L}Lh:~s_{\beta-\alpha}-\tan\beta c_{\beta-\alpha}=\sin{\alpha}/\cos{\beta},~~\bar{L}LH:~ c_{\beta-\alpha}+\tan\beta s_{\beta-\alpha}=\cos{\alpha}/\cos{\beta}
\ee
\be
\textnormal{Type I, III}:~~\bar{L}Lh:~ s_{\beta-\alpha}+\cot\beta c_{\beta-\alpha}=\cos{\alpha}/\sin{\beta},~~\bar{L}LH:~ c_{\beta-\alpha}-\cot\beta s_{\beta-\alpha}=\sin{\alpha}/\sin{\beta}
\label{lag}
\ee
Since the Higgs self couplings are independent of the type of the 2HDM, the production cross section in 2HDM types II and IV are the same. The same argument applies to types I and III, although, for a fixed value of $\alpha$ and large \tanb, type I and III couplings are smaller than the corresponding couplings in type II and IV. For a real analysis, one has to consider a final state which involves the charged Higgs decay to a lepton (or quark) pair such as $\tau\nu$ or $t\bar{b}$. From Tab. \ref{tabtypes}, it is concluded that a 2HDM type I is not relevant for a high \tanb regime since the charged Higgs decays are suppressed by a factor $\cot\beta$ and we are dealing with \tanb values as high as 10. Table \ref{tabtypes} indicates that 2HDM types II and IV are suitable for $H^{\pm}\ra \tau\nu$ as in other 2HDM types, this decay channel is suppressed by $\cot\beta$. Furthermore, 2HDM types II and III are suitable for $H^{\pm}\ra t\bar{b}$ which is suppressed in type IV. The choice of 2HDM type thus depends on the $\alpha$, $\beta$ and the charged Higgs decay channel. The $\alpha$ and $\beta$ parameters determine the production cross section and then one has to choose a proper 2HDM type for an analysis of a specific charged Higgs decay channel. Since type II and III are restricted by the lower limit on the charged Higgs mass at 300 GeV \cite{main2hdm}, and type I is not suitable for large \tanb, the following analysis which has a kinematic threshold at $m_{H^{\pm}}=250$ GeV for $\sqrt{s}=500$ GeV, is essentially relevant for a 2HDM Type IV.     
\section{Higgs self coupling and choices of $\alpha$ and $\beta$ parameters}   
The Higgs boson self-couplings for the vertices involved in the production of $\ell^+\ell^- \ra H^+H^-$ are expressed in Eqs. \ref{v1} and \ref{v2} \cite{2hdm3,2hdm4}.
\be
H^{\pm}H^{\pm}H:~\frac{-ie}{m_W\sin\theta_{W}\sin 2\beta}\left[(m_{H^{\pm}}^2-m_{A}^2+\frac{1}{2}m_{H}^2)\sin 2\beta\cos(\beta-\alpha)-(m_{H}^2-m_{A}^2)\cos 2\beta\sin(\beta-\alpha) \right] 
\label{v1}
\ee 
\be
H^{\pm}H^{\pm}h:~\frac{-ie}{m_W\sin\theta_{W}\sin 2\beta}\left[(m_{H^{\pm}}^2-m_{A}^2+\frac{1}{2}m_{h}^2)\sin 2\beta\sin(\beta-\alpha)+(m_{h}^2-m_{A}^2)\cos 2\beta\cos(\beta-\alpha) \right] 
\label{v2}
\ee 
Here two different regimes of large and small $\alpha$ are considered. The \tanb is set to 10 for numerical results which are obtained using COMPHEP 4.5.1 \cite{comphep,comphep2}. Two domains of $\alpha \simeq \beta \simeq \pi/2$, and $\beta \simeq \pi/2,~\alpha \simeq 0.1$ are explored. The $\alpha$ values below 0.1 turn out to be violating the vacuum potential constraints, as obtained by 2HDMC, therefore we restrict ourselves to $0.1<\alpha<\pi/2$.
\section{Different scenarios and results}
The aforementioned points in ($\alpha,\beta$) parameter space lead to the following simplification of the analysis. Two scenarios denoted as case I and II are described as follows.\\

\begin{figure}[h]
\begin{center}
\includegraphics[width=0.6\textwidth]{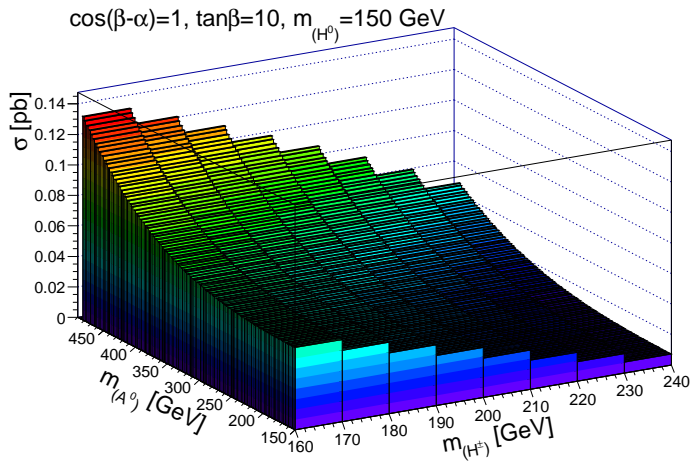}
\end{center}
\caption{The $\mu^+\mu^- \ra H^+H^-$ production cross section in terms of $m_{H^{\pm}}$ and $m_{A}$.}
\label{s1}
\end{figure}

\begin{figure}[h]
\begin{center}
\includegraphics[width=0.6\textwidth]{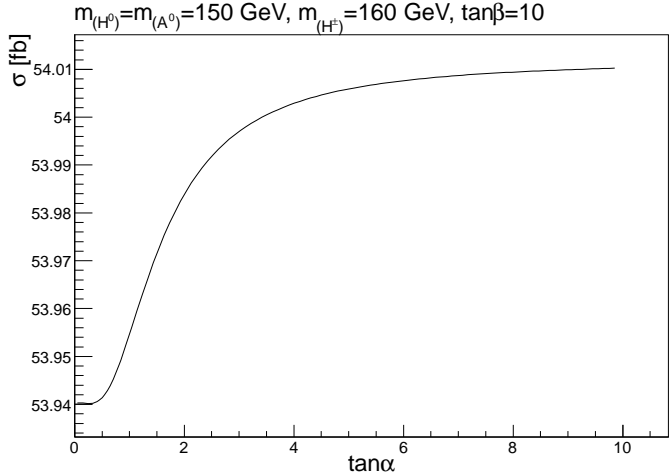}
\end{center}
\caption{The $\mu^+\mu^- \ra H^+H^-$ production cross section as a function of $\alpha$ parameter.}
\label{s2}
\end{figure}

\textbf{Case I}: $\alpha = \beta \simeq \pi/2$\\
With this setting, using Eqs. \ref{v1} and \ref{v2}, the Higgs self-couplings are dominated by $H^{\pm}H^{\pm}h$ which is effectively proportional to $(m_{h}^2-m_{A}^2)$. Since $m_{h}$ is set to 125 GeV to be consistent with LHC recent observations \cite{125atlas,125cms}, The two free parameters are then chosen as $m_A$ and $m_{H^{\pm}}$. This case can be expressed as $\cos(\beta-\alpha)=1,~\tan\beta=10$ as we have adopted \tanb = 10 as the example. Figure \ref{s1} shows the cross section of the charged Higgs pair production as a function of the charged Higgs and CP-odd Higgs masses. For all points, the value of $\Delta\rho$ parameter \cite{deltarho} is evaluated using 2HDMC and lies in the range $10^{-5}<\Delta\rho<10^{-3}$ which is acceptable by the experimental limits ($\Delta\rho<10^{-3}$ \cite{pdg}). As seen, the production cross section can be enhanced up to 130 $fb$ in the parameter space explored. Although we separated the two regimes of small and large $\alpha$, a scan over $\alpha$ values for a point in Fig. \ref{s1} (the front corner with lightest Higgs masses, $m_H=m_A=150$ GeV, $m_{H^{\pm}}=160$ GeV) shows that the cross section has a little dependence on the value of $\alpha$ parameter as illustrated in Fig. \ref{s2}.\\

\begin{figure}[h]
\begin{center}
\includegraphics[width=0.6\textwidth]{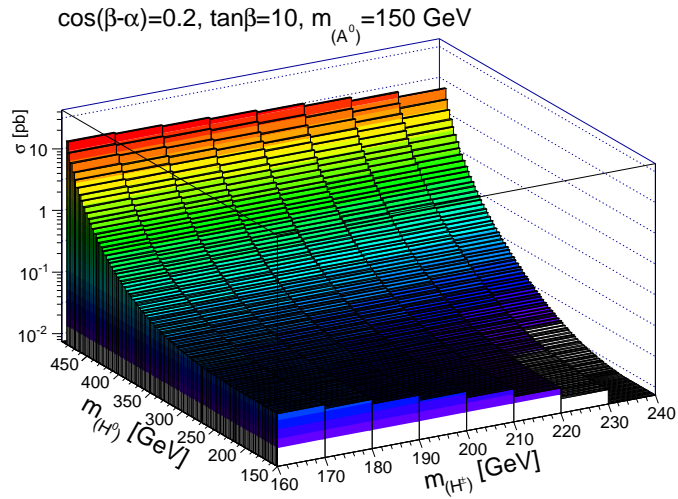}
\end{center}
\caption{The $\mu^+\mu^- \ra H^+H^-$ production cross section in terms of $m_{H^{\pm}}$ and $m_{H}$.}
\label{s3}
\end{figure}

\begin{figure}[h]
\begin{center}
\includegraphics[width=0.6\textwidth]{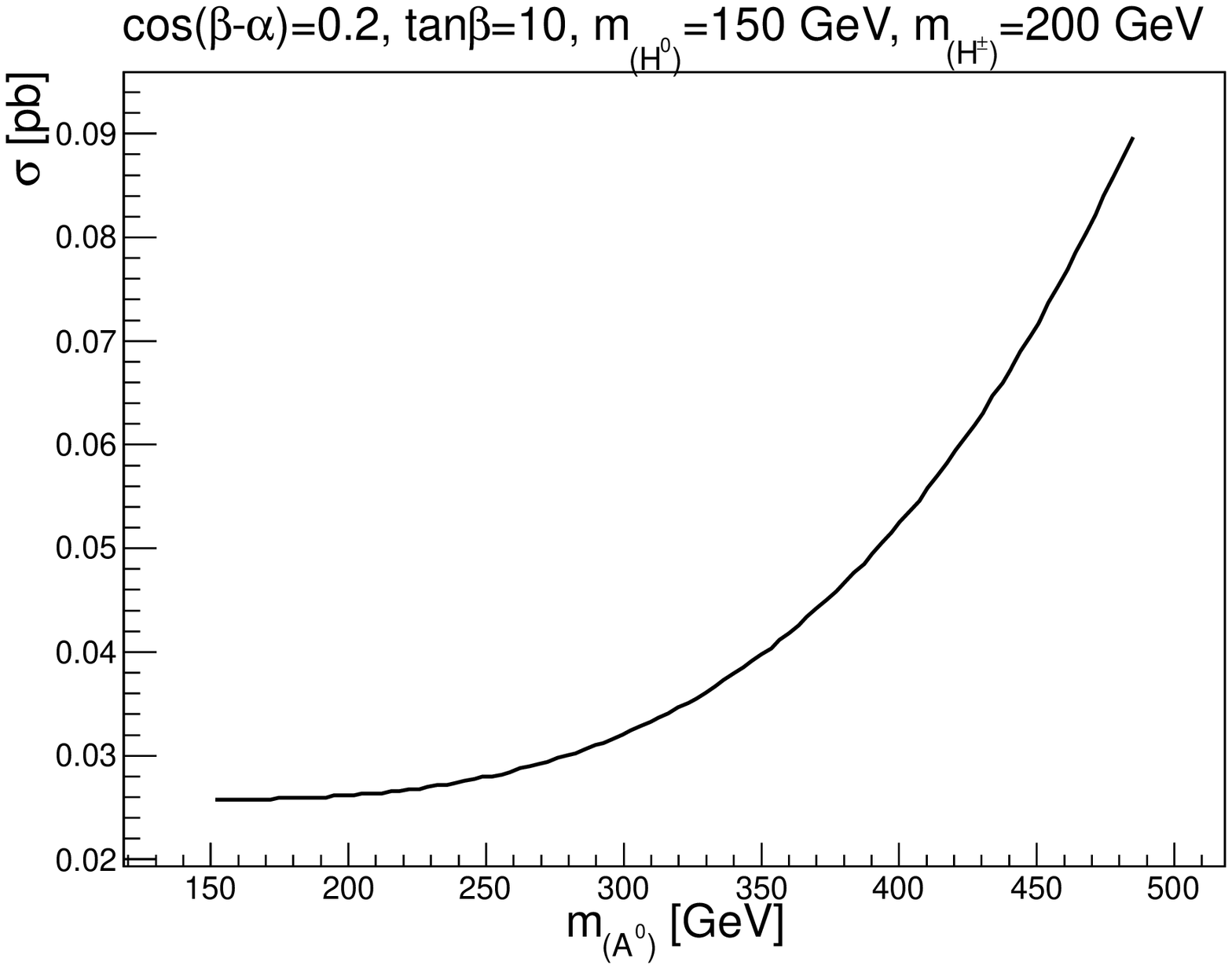}
\end{center}
\caption{The $\mu^+\mu^- \ra H^+H^-$ production cross section as a function of $m_{A}$ with $m_{H}=150$ GeV.}
\label{s4}
\end{figure}

\begin{figure}[h]
\begin{center}
\includegraphics[width=0.6\textwidth]{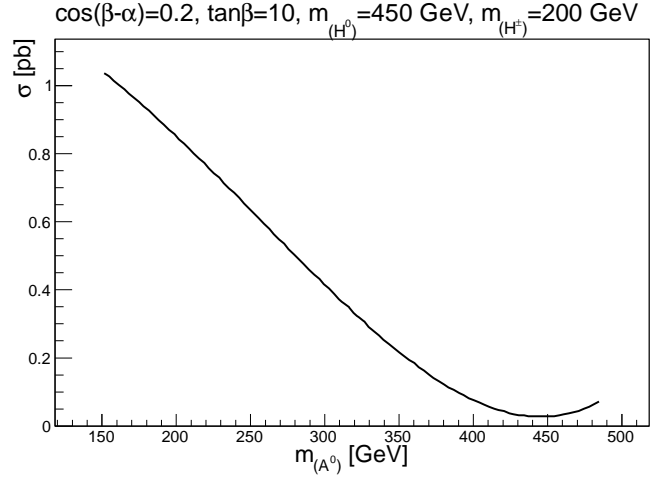}
\end{center}
\caption{The $\mu^+\mu^- \ra H^+H^-$ production cross section as a function of $m_{A}$ with $m_{H}=450$ GeV.}
\label{s55}
\end{figure}

\textbf{Case II}: $\alpha =0.1,~\beta \simeq \pi/2$:\\
Equations \ref{v1} and \ref{v2} show that with this setting, the Higgs self-couplings are dominated by $H^{\pm}H^{\pm}H$ which is proportional to $(m_{H}^2-m_{A}^2)$. Here, one can choose either $(m_A,m_{H^{\pm}})$ or $(m_H,m_{H^{\pm}})$ as the set of free parameters. However, the latter has a larger effect in cross section calculation because the CP-even Higgs, $H$, enters also in the propagator while the $A$ does not contribute. Therefore, in this case, the two free parameters are chosen as $m_H$ and $m_{H^{\pm}}$. This case can also be expressed as $\cos(\beta-\alpha)=0.2,~\tan\beta=10$. Figure \ref{s3} shows cross section values as a function of $m_H$ and $m_{H^{\pm}}$. All points satisfy $10^{-5}<\Delta\rho<10^{-4}$. As a conclusion, with small $\alpha$, the cross section shows a strong relation with $m_H$ and can be as high as 10 $pb$ with $m_H>450$ GeV. The variation of cross section with $m_A$ for a typical point, ($m_H=150$ GeV, $m_{H^{\pm}}=200$ GeV), is shown in Fig. \ref{s4} and confirms that there is less correlation with $m_A$ than with $m_H$. This point is not close to the neutral Higgs resonance, however, it was carefully selected to illustrate the effect of $m_{A}$ variation in the total cross section. With $m_{H^{\pm}}= 150$ GeV, which is the minimum value for $m_{A}$ in Fig. \ref{s4}, the cross section is essentially a quadratic function of $m_{A}$ which can be verified from Eq. \ref{v1}. If a heavier $H$ is chosen, the cross section does not necessarily increase with increasing $m_{A}$ due to the term $m_H^2-m_A^2$ which tends to zero for equal masses of the two Higgs bosons. In this case, starting with $m_{A}<m_{H}$, the cross section decreases with increasing $m_{A}$ until $m_{A}=m_{H}$, and then starts to grow quadratically. This feature has been shown in Fig. \ref{s55}. That is the reason a light $A$ was adopted in this case to avoid equality of $m_{A}$ and $m_{H}$, when a scan over $m_{H}$ is performed in the range $150~ \textnormal{GeV}~<~m_H~<~500~\textnormal{GeV}$.\\
Figure \ref{s3} shows that a sizable increase in the cross section is observed when increasing $m_{H}$. Moreover, for any fixed value of $m_{H}$, the cross section decreases monotonically with increasing the charged Higgs mass. A closer look at the general formula for the $s$-channel cross section, Eq. \ref{cseq} \cite{martinbook} shows that the cross section is proportional to the partial decay rate of the resonance (neutral Higgs) to the final state channel which is in turn related to the phase space factor and the square of the coupling of the neutral Higgs with final state particles involved in the process. In Eq. \ref{cseq}, $\Gamma_i(\Gamma_f)$ is the partial decay rate of the neutral Higgs to the incoming(outgoing) particles, $\sqrt{s}$ is the center of mass energy and $\Gamma$ is the total width of the neutral Higgs. The incoming particles are muons and outgoing particles are charged Higgs bosons. Therefore the relevant terms are the $H^+H^-H$ coupling, which is quoted in Eq. \ref{v1}, and the phase space factor. Inspecting Eq. \ref{v1} shows that the coupling increases with increasing $m_{H^{\pm}}$, however, the phase space factor involved in the decay rate decreases and tends to zero at $m_{H^{\pm}}\simeq \sqrt{s}/2=250$ GeV. Therefore two effects in Fig. \ref{cseq} are observed: the cross section enhancement when the resonance reaches the injected center of mass energy, i.e., $m_H\simeq \sqrt{s}=500$ GeV, and a rapid decrease of the cross section when the phase space saturation occures at $m_{H^{\pm}}\simeq \sqrt{s}/2=250$ GeV, however, no effect is observed at $m_{H^{\pm}}\simeq m_H/2$ as $m_H$ is off-shell and the relevant factor is $\sqrt{s}/2$.   
\begin{equation}
\sigma \sim \frac{\Gamma_i\Gamma_f}{(\sqrt{s}-m_{H})^2+\Gamma^2/4}
\label{cseq}
\end{equation}   


\section{Conclusion}
The charged Higgs pair production, $\ell^+\ell^- \ra H^+H^-$ is one of the main processes which would provide an observable signal in a wide range of the $m_{H^{\pm}},\tan\beta$ parameter space in MSSM. If a charged Higgs is observed at LHC in the coming years, this process would be the best candidate for a confirmation of LHC results at a linear collider with $\ell~=~e$. In this paper, it was shown that at a muon collider, i.e., with $\ell~=~\mu$, this process would be sensitive to the mass of neutral Higgs bosons which are involved in the s-channel Feynman diagrams. Two regimes of small and large $\alpha$ were adopted and it was concluded that with \tanb = 10 and $\cos(\beta-\alpha)=1$, the cross section is enhanced up to 130 $fb$ when $m_A$ is increased to the kinematic threshold $m_A=500$ GeV. On the other hand, with \tanb = 10 and $\cos(\beta-\alpha)=0.2$, the cross section can reach 10 $pb$ when $m_H$ is increased towards the kinematic threshold $m_H=500$ GeV. Since a charged Higgs pair production is observable with a cross section of 10 $fb$ or higher \cite{mine}, such large cross sections would provide observable signals earlier than expected from MSSM and therefore could be interpreted as a hint for a heavy neutral Higgs boson and the underlying theoretical model.

\end{document}